\documentclass[review]{elsarticle}
\usepackage{lineno}
\usepackage[colorlinks,
            linkcolor=blue,       
            anchorcolor=blue,  
            citecolor=blue,        
            ]{hyperref}
\usepackage{float} 
\usepackage{caption}
\usepackage[utf8]{inputenc}
\usepackage{makecell}
\usepackage{tikz}
\usepackage{amsfonts}
\usepackage{amsmath,amssymb}
\usepackage{systeme,mathtools}
\usetikzlibrary{shapes,snakes}
\tikzset{>=latex}
\DeclareMathAlphabet{\mathscr}{OT1}{pzc}{m}{it}
\usepackage{amssymb,amsmath}
\usepackage{geometry}
\usepackage{graphicx}
\usepackage{epstopdf}
\usepackage{booktabs}
\usepackage{color}
\usepackage{colortbl}
\usepackage{tabularx}
\usepackage{algorithmic}
\usepackage{listings}
\usepackage{booktabs} 
\usepackage{nomencl} 
\makenomenclature
\usepackage{adjustbox}
\usepackage{amsthm}                   
\newtheoremstyle{mystyle}
{3pt}    
{3pt}    
{}       
{}       
{\bfseries} 
{.}      
{.5em}   
{}       
\theoremstyle{mystyle}             
\newtheorem{definition}{Definition}

\newtheorem{theorem}{Theorem}
\newtheorem{example}{Example}

\usepackage{multirow}
\usepackage{graphicx}
\usepackage{subfigure}
\usepackage{tikz}
\usepackage{threeparttable}
\usepackage[ruled]{algorithm2e}
\usepackage{bm}

\begin{document}
\begin{frontmatter}
\title{Grey graphs and its application}
\author[label1]{Wanli Xie }
\author[label1]{Jiale Zhang\corref{cor1}}
\author[label1]{Ruiqing Cao}
\address[label1]{School of Communication, Qufu Normal University, Rizhao, China}
\cortext[cor1]{Corresponding author: Jiale Zhang (Jiale\_Z@qfnu.edu.cn).}
\begin{abstract}
In multi-attribute decision-making problems where the attribute values are interval grey numbers, a simplified form based on kernels and the degree of greyness is presented. Combining fuzzy graph theory with the kernel and the degree of greyness of interval grey numbers, grey graphs and their corresponding operation rules are presented. This paper presents a new multi-attribute decision-making method based on grey graph theory. We analyzed and evaluated the alternative schemes using grey graph. Lastly, a numerical example was conducted in order to demonstrate the effectiveness and feasibility of the proposed method.
\end{abstract}
\begin{keyword}
Multi-attribute decision making\sep Grey graph\sep Interval grey numbers \sep Grey system
\end{keyword}
\end{frontmatter}
\section{Introduction}
A key component of modern decision theory is the multi-attribute decision analysis. Under the condition of multiple attributes, it focuses on selecting limited candidate schemes. During actual operations, the decision-making information that experts face often contains a wide range of uncertainties and fuzziness. The fuzzy set concept was proposed by Zadeh as a solution to this problem, which is capable of representing both positive and negative states with just one membership degree \cite{zadeh1965fuzzy}. It extends the definition of membership degree in the original set from 0 and 1 to any real number between 0 and 1, so as to achieve a more precise understanding of the fuzzy concept. Researchers were intrigued by Zadeh's ideas. Atanassov developed intuitionistic fuzzy sets (IFS) in 1986, expanding the theory of fuzzy sets \cite{atanassov2016intuitionistic}. As well as maintaining the original membership degree, IFS also incorporate the concepts of non-membership degree and hesitation degree. Torra subsequently proposed the concept of hesitant fuzzy sets (HFS) \cite{torra2010hesitant}. With this concept, an element's membership degree is no longer a single definite value, but rather a set of possible values within the interval [0,1], which is more objectively indicative of the uncertainty and hesitation of decision-makers during the decision-making process. The fuzzy graph theory was proposed by Rosenfeld and it introduced the concept and related properties of fuzzy graphs \cite{rosenfeld1975fuzzy}. Later, the fuzzy graph theory was developed and improved in the context of multi-attribute decision-making. Zhang incorporated hesitant fuzzy sets and fuzzy graphs, proposed hesitant fuzzy graphs, and constructed a multi-attribute decision-making framework suitable for hesitant fuzzy environments based on these concepts \cite{zhang2017hesitant}. Feng developed a new decision-making algorithm based on intuitionistic hesitant fuzzy sets by combining their characteristics with fuzzy graph theory \cite{JSJZ202502053}. All of these methods are capable of handling the complex interactions among multiple attributes of alternative solutions to deal with difficult decision-making problems involving uncertainty and fuzziness.

In these decision-making models, however, the decision parameters are usually precise numerical values. In actual decision-making scenarios, nevertheless, due to the limited professional knowledge of experts, the tight schedule, as well as individual differences in cognition and the complexity of decision-makers' thinking, they are often unable to provide specific numerical values as effect measures or indicator weights. In the absence of specific numerical values, decision-makers are only able to provide interval numbers to reflect their views (the lower limit represents the most conservative view, whereas the upper limit indicates the most optimistic view). Practically, this interval range reflects the uncertainty and fuzzy nature of decision-making problems, thus precluding decision-makers from using deterministic decision-making methods. Decision-making problems involving uncertainty are prevalent in multiple fields such as society \cite{qi2024fermatean}, technology \cite{zhang2024entropy}, environment \cite{zhang2024multi}, and management \cite{zia2024complex}. Uncertain decision-making methods are not only of theoretical significance, but they also have practical application. In 1982, Deng proposed the grey system theory, which aims to whiten grey areas through inference and exploration of limited known information and modeling with limited datasets \cite{Deng1982control}. This theory can be used to construct more flexible decision-making models for uncertain multi-attribute problems as shown in Table \ref{tab:addlabel} in recent years. With such a model, the decision-making results are more accurate, so they are more adaptable and effective. The interval numbers were first proposed by Young \cite{young1931algebra}. In general, they can be any number within a specified range. To reflect the complexity and uncertainty of the objective world, decision-making information frequently uses grey numbers rather than interval numbers. An approximate range of a grey number is known, but its exact value is uncertain \cite{julong1989introduction}. Interval grey numbers is theoretically based on grey system. It is used to describe numbers whose values are uncertain within a specific range. As a result of their inherent complexity, such numbers cannot be fully represented by one exact real value. The interval grey numbers not only cover the entire range of uncertain information, but also reveal data with unclear greyness within the interval. Regarding interval grey number operations, Fang proposed the concept of the first and second standard interval grey numbers as well as conversion rules between ordinary interval grey numbers and standard interval grey numbers \cite{ZHYJ200400001033}. Using the standardized definition of greyness \cite{sifeng2006measures}, Liu introduces the concept of greyness to handle grey intervals generated by interval grey number operations \cite{liu2010algorithm}. In addition, he proposed the definition of grey numbers’ kernel. Following this, he constructed the operating rules, which simplified grey number operations into real number operations based on the greyness of grey numbers. According to Guo \cite{guo2016multi}, a multi-attribute decision-making problems with interval grey numbers and weight information can be based on the kernel and greyness of the interval grey numbers.
\begin{table}[htbp]
	\centering
	\caption{Recent studies of decision-making methods for uncertain multi-attribute problems used grey theory.}
	  \begin{adjustbox}{max width=\textwidth}
	\begin{tabular}{llp{14.585em}lp{25.125em}}
		\toprule
		Field & Reference & Objective & \multicolumn{1}{p{10.375em}}{Method} & Main contribution \\
		\midrule
		\multirow{4}[1]{*}{Society} & Xu et al. \cite{xu2021evaluation}   & Smart community elderly care service quality analysis & \multicolumn{1}{p{10.375em}}{IVIFE} & It promotes service standardization and reduces regional differences in smart community elderly care services. \\
		& Zhang et al.  \cite{zhang2022large} &Extract the key factors influencing pandemic control & \multicolumn{1}{p{10.375em}}{DEMATEL} & The interdependence of relationships and system components can be examined using the DEMATEL framework. \\
		& Qi et al. \cite{qi2024fermatean}    & The selection of electric vehicle charging stations & \multicolumn{1}{p{10.375em}}{CFFRS, CIVFFRS} & CFFRS and CIVFFRS models indicate that Jamuria would be the most suitable location to build an electric vehicle charging station. \\
		& Zhang et al. \cite{zhang2024dynamic}  & An emergency of public health & \multicolumn{1}{p{10.375em}}{Quantum-like Bayesian networks} & An innovative approach to addressing cognitive biases and information fusion is presented in this paper. \\
		&       & \multicolumn{1}{l}{} &       & \multicolumn{1}{l}{} \\
		\multirow{4}[0]{*}{Technology} & Liu et al. \cite{liu2022solving}  & C919's key component selection & \multicolumn{1}{p{10.375em}}{The group of weight vector with kernel} & Grey clustering evaluation can be solved effectively by a novel two-stage decision model based on a group of weight vectors with kernels and a weighted comprehensive clustering coefficient vector. \\
		& James et al. \cite{james2023purchase}& Choose the right equipment for automobile maintenance garages & \multicolumn{1}{p{10.375em}}{Fuzzy AHP, GRA} & In this way, according to the garage management's purchase criteria, each supplier of equipment can be prioritized. \\
		& Li and Li \cite{li2024grey}   & An evaluation of the technological innovation capability & \multicolumn{1}{p{10.375em}}{Generalized greyness of interval grey number} & It is appropriate for decision-making problems involving interval grey numbers and real numbers. \\
		& Zhang and Li \cite{zhang2024entropy}& Selecting a brackish water irrigation pattern for winter wheat & \multicolumn{1}{p{10.375em}}{TOPSIS} & It provides a new method for determining unknown attribution weights and values using interval grey numbers, which coexist with real numbers. \\
		&       & \multicolumn{1}{l}{} &       & \multicolumn{1}{l}{} \\
		\multirow{4}[0]{*}{Environment} & Tan et al. \cite{tan2022decision}  & Assessment of typhoon disaster & \multicolumn{1}{p{10.375em}}{Single-value neutrosophic set, GRA} & The results of such assessments can provide intelligent decision support to the relevant disaster management agencies. \\
		& Zheng et al. \cite{zheng2024grey}& Aviation ecological assessment & \multicolumn{1}{p{10.375em}}{Dual hesitant fuzzy set} & Based on prospect theory in a dual hesitant fuzzy environment, a model of grey target decision-making is developed. \\
		& Guo et al. \cite{guo2024grey}  & Selection of low-carbon suppliers & \multicolumn{1}{p{10.375em}}{IGN, SNA, CPR} & GMCGCDM is capable of expanding the calculation method for DM's weights and consensus degrees. \\
		& Zhang et al. \cite{zhang2024multi}& Planning for pumped storage capacity & \multicolumn{1}{p{10.375em}}{Multi-attribute decision-making method} & A capacity planning problem for pumped storage stations in hybrid operation systems is considered in this paper. \\
		&       & \multicolumn{1}{l}{} &       & \multicolumn{1}{l}{} \\
		\multirow{4}[1]{*}{Management} & Xu and Yang \cite{xu2022service}   & Assessment of medical care and nursing institutions & \multicolumn{1}{p{10.375em}}{IVPFS-DEMATEL} & The proposed methodology is effective in assessing the quality of elderly care. \\
		& Huang et al. \cite{huang2022vetoed}& Official vehicle supplier selection & \multicolumn{1}{p{10.375em}}{Multi-objective grey target, Veto function} & It proposes a multi-attribute vetoed grey target decision method. \\
		& Asnaashari et al. \cite{asnaashari2023impact}& Selecting contractors based on claim management & OPA-G & Based on this method, the criterion "Hiring a technical team with experience and education" has the first in selecting contractors. \\
		& Zia et al. \cite{zia2024complex}  & Selection of warehouse distributors and evaluation of faculty candidates & CLDFS & CLDFS is an advanced framework for dealing with ambiguity when periodicity is present. \\
		\bottomrule
	\end{tabular}%
	  \end{adjustbox}
	\label{tab:addlabel}%
\end{table}%

Combining the kernel and greyness of interval grey numbers, as well as the application advantages of fuzzy graphs in complex scenarios of multi-attribute decision-making, especially in describing the interaction between attributes for decision-making outcomes. This study combines fuzzy graph theory with grey numbers, introduces grey graphs (GG), and establishes corresponding operation rules. Using grey graph theory, this study develops a novel multi-attribute decision-making method. Grey graphs are used in this method to analyze and evaluate alternative schemes for fuzzy environments in order to address interaction issues among attributes. As a result, the proposed method was tested on specific cases to make sure it was effective and practicable.

The paper is structured as follows: Section 2 offers a brief overview of fuzzy graphs and grey numbers, as well as presents the concept of grey graphs. Section 3 provides a description of the modeling process for the interval grey number multi-attribute decision-making method based on grey graphs. Through numerical examples, section 4 evaluates the effectiveness and feasibility of the proposed method. Section 5 presents conclusions and perspectives.
\section{Preliminaries}
\subsection{Fuzzy Graph Theory}
\begin{definition} [\cite{zadeh1965fuzzy}]
Suppose $X$ is a space of objects, and $x$ indicates the generic element of $X$. There is a fuzzy set $A$ in $X$ characterized by a membership function ${\mu _{\rm{A}}}$ which is associated with each object in X a real number from 0 to 1, with ${\mu _{\rm{A}}}(x)$ representing $x$'s grade of membership in $A$, expressed as
\[A = \{ (x,{\mu _A}(x))|x \in X\}. \]
\end{definition}
Assume that there are two fuzzy sets $A$ and $B$ on $X$, as well as ${A^C}$ represents the complement of $A$. The subsequent equation is valid
\[\begin{array}{l}(1)A = B \Leftrightarrow \forall {\rm{x}} \in X,\ {\mu _A}(x) = {\mu _B}(x);\\
(2)A \subset B \Leftrightarrow \forall {\rm{x}} \in X,\ {\mu _A}(x) \le {\mu _B}(x);\\
(3){A^C} = \{ (x,1 - {\mu _A}(x))|x \in X\};\\
(4)A \cup B = \{ (x,{\mu _A}(x) \vee {\mu _B}(x))|x \in X\}  = \max [{\mu _A}(x),{\mu _B}(x)];\\
(5)A \cap B = \{ (x,{\mu _A}(x) \wedge {\mu _B}(x))|x \in X\}  = \min [{\mu _A}(x),{\mu _B}(x)].\end{array}\]
\begin{definition} [\cite{rosenfeld1975fuzzy}]
A fuzzy graph is described as $G = (V,E,\sigma ,\mu )$. In this scenario, $\mu$ is the symmetric fuzzy relation on $\sigma$. $V$ represents the set of vertices, $E$ represents the set of edges, $\sigma$ represents the fuzzy set of vertices on $V$, as well as $\mu$ represents the fuzzy set of edges on E. It exists \[\sigma :V \to [0,1],\ \mu :V \times V \to [0,1].\] For all $u \in V,\ v \in V$, we have $\mu (uv) \le \sigma (u) \wedge \sigma (v)$, $uv$ is the edge between two vertices $u$ and $v$.
\end{definition}
\subsection{The Kernel and Degree of Greyness of Interval Grey Numbers}
	\begin{definition} [\cite{liu2010algorithm}]
Let $\Omega = [a, \overline{a}]$ be an interval number. A \emph{grey number} $\otimes \in [a, \overline{a}]$ can be expressed in simplified form as 
		\[
		\hat{\otimes}(g_0) = \left( \frac{a + \overline{a}}{2}, \; \frac{\overline{a} - a}{\mu(\Omega)} \right),
		\] where $\mu(\Omega)$ denotes the measure of the domain $\Omega$. When $\Omega$ is normalized to $[0,1]$, it follows that $g_0 = \overline{a} - a$. Here, the first component $\displaystyle \hat{\otimes} = \frac{a + \overline{a}}{2}$ is called the \emph{kernel}, and the second component $g_0$ is called the \emph{grey degree}. This definition maps an interval grey number bijectively onto $(\hat{\otimes}, g_0)$, which facilitates computation while retaining all information.
	\end{definition}
	\begin{theorem} [\cite{liu2010algorithm}]
		For the operations of addition, subtraction, multiplication, and division among grey numbers, the resulting grey degree shall not be less than the maximum grey degree of the operands. Hence, all grey number operations satisfy
		\[\begin{array}{l}(1)(\hat x,g_x)\oplus(\hat y,g_y) = (\hat{x}, g_x) + (\hat{y}, g_y)=(\hat x+\hat y,\, g_x\vee g_y);\\
		(2)(\hat x,g_x)\otimes(\hat y,g_y) = (\hat{x}, g_x) \times (\hat{y}, g_y)=(\hat x\hat y,\, g_x\vee g_y); \\
		(3)c\odot(\hat x,g_x)=c \times (\hat{x}, g_x) = (c\hat x,\,g_x),\ c \in \mathbb{R}.\end{array}\]
		That is, the kernel follows ordinary arithmetic rules, while the grey degree takes the maximum value among participants.	
	\end{theorem}
	\begin{theorem} [\cite{yan2014ranking}]
		Let $(\hat{\otimes}, g_0)$ be a standard grey number, \[
		\delta(\otimes) = \gamma(\otimes) \cdot \hat{\otimes},
		\quad \text{where} \quad
		\gamma(\otimes) = \frac{1}{1 + g_0}.
		\] Then $\delta(\otimes)$ is the relative kernel of the grey number, and $\gamma(\otimes)$ reflects the precision of the grey number (a smaller grey degree implies higher precision). To compare two grey numbers, one should first compare their relative kernels $\delta(\otimes)$. The larger kernel, the greater grey number. If the kernels are equal, the comparison is then determined by $\gamma(\otimes)$, where a smaller grey degree is preferred.
	\end{theorem}
\subsection{Grey Graphs}
	\begin{definition}
We propose a grey graph $G=(V,E,\sigma,\mu)$. $\sigma$ represents the grey number set of vertices on $V$, as well as $\mu$ represents the grey number set of edges on $E$. $\sigma$ and $\mu$ are assigned a \emph{standard grey number} $\tilde x=(\hat x,g_x)$, where $\hat x$ is the kernel and $g_x$ the degree of greyness. The resulting grey graph satisfies this condition $\mu (pq) \le \min (\sigma (p),\sigma (q)),$ $\forall pq \in E.$ This means edges are not stronger than their endpoints, which edges exhibit greater uncertainty (smaller $\hat x$, larger $g_x$) than their incident vertices. 
	\end{definition}
	\begin{example}
		Let $G=(V,E,\sigma,\mu)$ be a grey graph. $V = \{ {x_1},{x_2},{x_3},{x_4},{x_5}\},\ E = \{ {x_1}{x_2},{x_1}{x_3},{x_1}{x_4},{x_1}{x_5},\\{x_2}{x_3},{x_2}{x_4},{x_2}{x_5},{x_3}{x_4},{x_3}{x_5},{x_4}{x_5}\}.$ $\sigma$ is the grey number set on $V$, and $\mu$ is the grey number set on $E$. $
			\sigma  = \{ ({x_1},(0.7,0.2)),({x_2},(0.6,0.1)),({x_3},(0.9,0.3)),({x_4},(0.8,0.5)),({x_5},(0.5,0.4))\} ,$ $\mu  = \{ ({x_1}{x_2},(0.5,0.3)),\\({x_1}{x_3},(0.6,0.4)),({x_1}{x_4},(0.5,0.6)),({x_1}{x_5},(0.4,0.8)),({x_2}{x_3},(0.4,0.4)),({x_2}{x_4},(0.2,0.7)),({x_2}{x_5},(0.1,0.5)),\\({x_3}{x_4},(0.3,0.6)),({x_3}{x_5},(0.4,0.5)),({x_4}{x_5},(0.1,0.8))\}.$ The first data is kernel, the second is grey degree. The grey graph is shown in Figure \ref{fig:1}.
			\begin{figure}[H]
				\centering
				\includegraphics[width=0.4\linewidth]{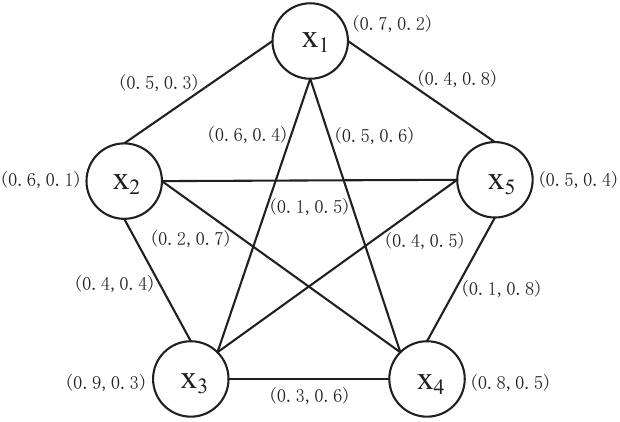}
				\caption{ Grey graph.}
				\label{fig:1}
			\end{figure}
		\end{example}
\begin{definition}
		Set $G=(V,E,\sigma,\mu)$ be a grey graph. There are two vertices $p$ and $q$ on $V$, for all $pq$ $\in$ $E$. If
		\[\mu (pq) = \min (\sigma (p),\sigma (q)),\ \forall pq \in E,\]
		\[\hat  \otimes_ {\mu }(pq) = \min (	\hat  \otimes_ {\sigma} ( p),\hat  \otimes_ {\sigma }( q)),\ {g_0}_{\mu }(pq) = \max ({g_0}_{\sigma }({p}),{g_0}_{\sigma} (q)),\]
		the resulting grey graph satisfies the strong grey graph.
\end{definition}
\begin{example}
 Here is a grey graph $G=(V,E,\sigma,\mu)$.	$V = \{ {x_1},{x_2},{x_3},{x_4}\},\ E = \{ {x_1}{x_2},{x_1}{x_3},{x_1}{x_4},{x_2}{x_3},{x_2}{x_4},\\{x_3}{x_4}\},$\ $
	\sigma  = \{ ({x_1},(0.5,0.6)),({x_2},(0.3,0.5)),({x_3},(0.7,0.2)),	({x_4},(0.4,0.7))\} .$
Based on strong grey graph condition, $\mu_{{x_1}{x_2}}=(\min (	\hat  \otimes_ {\sigma} ( x_1),\hat  \otimes_ {\sigma }( x_2)),\max ({g_0}_{\sigma }({x_1}),{g_0}_{\sigma} (x_2))) =(0.3,0.6),$ we could compute $\mu  = \{ ({x_1}{x_2},(0.3,0.6)),({x_1}{x_3},(0.5,0.6)),({x_1}{x_4},(0.4,0.7)),({x_2}{x_3},(0.3,0.5)),({x_2}{x_4},(0.3,0.7)),({x_3}{x_4},(0.4,0.7))\}.$ The strong grey graph is introduced in Figure \ref{fig:2}.
	\begin{figure}[H]
		\centering
		\includegraphics[width=0.4\linewidth]{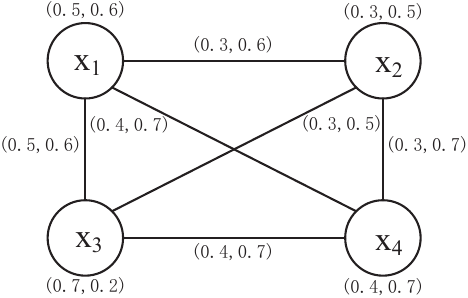}
		\caption{Strong grey graph.}
		\label{fig:2}
	\end{figure}
\end{example}
\begin{definition}
There are two grey graphs $G_1=(V_1,E_1,\sigma_1,\mu_1),$ $G_2=(V_2,E_2,\sigma_2,\mu_2).$

(1) The union of $G_1$ and $G_2$ is ${G_1} \cup {G_2} = ({\sigma _1} \cup {\sigma _2},{\mu _1} \cup {\mu _2}).$ It has
\[\begin{array}{l}
\textbf{a.}\ \hat  \otimes_ {{\sigma _1} \cup {\sigma _2}}(p) = \hat  \otimes_{{\sigma _1}}(p),\ {g_0}_{{\sigma _1} \cup {\sigma _2}}(p) = {g_0}_{{\sigma _1}}(p),\ \forall p \in {V_1},\forall p \notin {V_2};\\
	\textbf{b.}\ \hat  \otimes_ {{\sigma _1} \cup {\sigma _2}}(p) = \hat  \otimes_{{\sigma _2}}(p),\ {g_0}_{{\sigma _1} \cup {\sigma _2}}(p) = {g_0}_{{\sigma _2}}(p),\ \forall p \notin {V_1},\ \forall p \in {V_2};\\
	\textbf{c.}\ \hat  \otimes_ {{\sigma _1} \cup {\sigma _2}}(p) = \max (\hat  \otimes_ {{\sigma _1}}(p),\hat  \otimes_ {{\sigma _2}}(p)),\ {g_0}_{{\sigma _1} \cup {\sigma _2}}(p)=\min ({g_0}_ {{\sigma _1}}(p),{g_0}_{{\sigma _2}}(p)),\ \forall p \in {V_1} \cap {V_2};\\
\textbf{d.}\ 	\hat  \otimes_ {{\mu _1} \cup {\mu _2}}(pq) = \hat  \otimes_{{\mu _1}}(pq),\ {g_0}_{{\mu _1} \cup {\mu _2}}(pq) = {g_0}_{{\mu _1}}(pq),\ \forall pq \in {E_1},\ \forall pq \notin {E_2};\\
	\textbf{e.}\ 	\hat  \otimes_ {{\mu _1} \cup {\mu _2}}(pq) = \hat  \otimes_{{\mu _2}}(pq),\ {g_0}_{{\mu _1} \cup {\mu _2}}(pq) = {g_0}_{{\mu _2}}(pq),\ \forall pq \notin {E_1},\ \forall pq \in {E_2};\\
\textbf{f.}\ 	\hat  \otimes_ {{\mu _1} \cup {\mu _2}}(pq) = \max (\hat  \otimes_ {{\mu _1}}(pq),\hat  \otimes_ {{\mu _2}}(pq)),\ {g_0}_{{\mu _1} \cup {\mu _2}}(pq) = \min ({g_0}_ {{\mu _1}}(pq),{g_0}_ {{\mu _2}}(pq)),\ \forall pq \in {E_1} \cap {E_2}.
\end{array}\]

(2) The addition of $G_1$ and $G_2$ is ${G_1} + {G_2} = ({\sigma _1} + {\sigma _2},{\mu _1} + {\mu _2}).$ It has
\[\begin{array}{l}
\textbf{a.}\ \hat  \otimes_ {{\sigma _1} + {\sigma _2}}(p) = \hat  \otimes_ {{\sigma _1} \cup {\sigma _2}}(p),\ {g_0}_{{\sigma _1} + {\sigma _2}}(p) = {g_0}_{{\sigma _1} \cup {\sigma _2}}(p),\ \forall p \in {V_1} \cup {V_2};\\
\textbf{b.}\ \hat  \otimes_ {{\mu _1} + {\mu _2}}(pq) = \hat  \otimes_ {{\mu _1} \cup {\mu _2}}(pq),\ {g_0}_{{\mu _1} + {\mu _2}}(pq) = {g_0}_{{\mu _1} \cup {\mu _2}}(pq),\ \forall pq \in {E_1} \cup {E_2};\\
\textbf{c.}\ \hat  \otimes_ {{\mu _1} + {\mu _2}}(pq) = \min (\hat  \otimes_ {\mu _1}(pq),\hat  \otimes_ {\mu _2}(pq)),{g_0}_{{\mu _1} + {\mu _2}}(pq) = \max ({g_0}_{\mu _1}(pq),{g_0}_{\mu _2}(pq)),\ \forall pq \in E.
\end{array}\]

(3) The Cartesian product of $G_1$ and $G_2$ is ${G_1} \times {G_2} = ({\sigma _1} \times {\sigma _2},{\mu _1} \times {\mu _2}).$ It has
\[\begin{array}{l}

\textbf{a.}\ \hat  \otimes_ {{\sigma _1} \times {\sigma _2}}(p,q) = \min (\hat  \otimes_ {{\sigma _1}}(p),\hat  \otimes_ {{\sigma _2}}(q)),\\\ \ \ \ {g_0}_{{\sigma _1} \times {\sigma _2}}(p,q)=\max ({g_0}_ {{\sigma _1}}(p),{g_0}_{{\sigma _2}}(q)),\ \forall p,q \in {V_1} \times {V_2};\\
\textbf{b.}\ \hat  \otimes_ {{\mu _1} \times {\mu _2}}((r,p),(r,q)) = \min (\hat  \otimes_ {{\mu _1}}(r),\hat  \otimes_ {{\mu _2}}(pq)),\\\ \ \ \ {g_0}_{{\mu _1} \times {\mu _2}}((r,p),(r,q)) = \max ({g_0}_ {{\mu _1}}(r),{g_0}_ {{\mu _2}}(pq)),\ \forall r \in {V_1},\ pq \in {E_2};\\
	\textbf{c.}\ \hat  \otimes_ {{\mu _1} \times {\mu _2}}((p,r),(q,r)) = \min (\hat  \otimes_ {{\mu _1}}(pq),\hat  \otimes_ {{\mu _2}}(r)),\\\ \ \ \ {g_0}_{{\mu _1} \times {\mu _2}}((p,r),(q,r)) = \max ({g_0}_ {{\mu _1}}(pq),{g_0}_ {{\mu _2}}(r)),\ \forall r \in {V_2},\ pq \in {E_1}.
\end{array}\]
\end{definition}
\section{An Interval Grey Numbers Multi-Attribute Decision-Making Model Based on Grey Graph}
\subsection{Problem Formulation}
Consider an uncertain multi-attribute decision-making problem consisting of $n$ alternative schemes and $m$ evaluation criteria. Let $X=\{X_1,\dots,X_n\}$ be the set of alternatives and $A=\{A_1,\dots,A_m\}$ the attributes. Each attribute $A_j$ is assigned a weight $\omega_j$, forming the attribute weight vector $\boldsymbol{\omega} = (\omega_1, \omega_2, \ldots, \omega_m),$ where $\omega_j^L \leq \omega_j \leq \omega_j^U, 0 \leq \omega_j^L \leq \omega_j^U \leq 1,$ and $\sum_{j=1}^{m} \omega_j = 1.$ The decision matrix R can be formed if the decision-maker provides decision-making information in grey numbers for the alternative option X under attribute A, denoted as $(\hat r_{ij},g_{r_{ij}})$ with interval-to-$[0,1]$ normalization. From this, a grey graph $G=(V,E,\sigma,\mu)$ that describes the interrelationships of attributes is established. Each attribute vertex $A_j$ carries its own grey weight $\sigma_{A_j}=(\hat w_j,g_{w_j})$. To capture inter-attribute interactions we introduce an attribute-layer influence network with edges $(A_p,A_q)$ and its grey weights $\mu_{pq}=(\hat\xi_{pq},g_{\xi_{pq}})$.
\subsection{Construction of the Evaluation Matrix}
For each alternative $X_i$ and attribute $A_j$, the evaluation value is represented as an interval grey number \[z_{ij} = [z_{ij}^L, z_{ij}^U], \quad i = 1,2,\ldots,n; \; j = 1,2,\ldots,m.
\] The lower and upper bounds, $z_{ij}^L$ and $z_{ij}^U$, respectively represent the minimum and maximum possible evaluation values. The decision matrix of all alternatives with respect to all criteria can then be written as \[Z = \begin{bmatrix}
	[z_{11}^L, z_{11}^U] & [z_{12}^L, z_{12}^U] & \cdots & [z_{1m}^L, z_{1m}^U] \\
	[z_{21}^L, z_{21}^U] & [z_{22}^L, z_{22}^U] & \cdots & [z_{2m}^L, z_{2m}^U] \\
	\vdots & \vdots & \ddots & \vdots \\
	[z_{n1}^L, z_{n1}^U] & [z_{n2}^L, z_{n2}^U] & \cdots & [z_{nm}^L, z_{nm}^U]
\end{bmatrix}.
\]
\subsection{Normalization of the Decision Matrix}
To ensure comparability among different attributes, the decision matrix must be normalized. For each attribute $A_j$, define \[
z_j^- = \min_i (z_{ij}^L), \quad
z_j^+ = \max_i (z_{ij}^U), \quad
d_j = z_j^+ - z_j^-,
\] where $d_j$ is the range of attribute $A_j$.

For benefit-type attributes (where larger values are preferred), normalization is performed as
\begin{align}
	r_{ij}^L &= \frac{z_{ij}^L - z_j^-}{d_j}, \\
	r_{ij}^U &= \frac{z_{ij}^U - z_j^-}{d_j}.
\end{align}
For cost-type attributes (where smaller values are preferred), normalization is defined as
\begin{align}
	r_{ij}^L &= \frac{z_j^+ - z_{ij}^U}{d_j}, \\
	r_{ij}^U &= \frac{z_j^+ - z_{ij}^L}{d_j}.
\end{align}
The normalized decision matrix is then expressed as
\[
R = 
\begin{bmatrix}
	[r_{11}^L, r_{11}^U] & [r_{12}^L, r_{12}^U] & \cdots & [r_{1m}^L, r_{1m}^U] \\
	[r_{21}^L, r_{21}^U] & [r_{22}^L, r_{22}^U] & \cdots & [r_{2m}^L, r_{2m}^U] \\
	\vdots & \vdots & \ddots & \vdots \\
	[r_{n1}^L, r_{n1}^U] & [r_{n2}^L, r_{n2}^U] & \cdots & [r_{nm}^L, r_{nm}^U]
\end{bmatrix}.
\]
\subsection{Simplified Form Representation}
The normalized matrix $R$ can be further transformed into a simplified form based on kernel and greyness \[
R =
\begin{bmatrix}
	\otimes_{11}(g_{11}^0) & \otimes_{12}(g_{12}^0) & \cdots & \otimes_{1m}(g_{1m}^0) \\
	\otimes_{21}(g_{21}^0) & \otimes_{22}(g_{22}^0) & \cdots & \otimes_{2m}(g_{2m}^0) \\
	\vdots & \vdots & \ddots & \vdots \\
	\otimes_{n1}(g_{n1}^0) & \otimes_{n2}(g_{n2}^0) & \cdots & \otimes_{nm}(g_{nm}^0)
\end{bmatrix}.
\]
\subsection{Decision-Making Procedure}
\begin{figure}[H]
		\centering
		\includegraphics[width=1\linewidth]{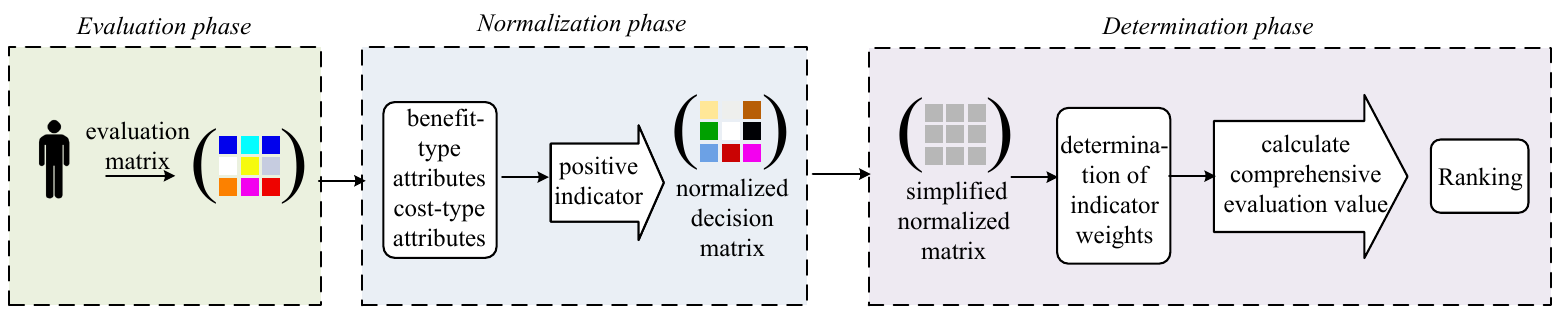}
		\caption{ Schematic diagram of proposed GG method.}
		\label{fig:3}
	\end{figure}

The corresponding schematic diagram of the decision making procedure is shown in Figure \ref{fig:3}. The overall decision-making process for the interval grey numbers multi-attribute model based on grey graph can be summarized as follows:
\\
\textbf{Step 1:} Construct the interval evaluation matrix $Z$ based on the decision problem and available data.\\
\textbf{Step 2:} Compute the domain range $d_j$ for each attribute, normalize all attributes according to their type, and obtain the simplified normalized matrix $R$.\\
\textbf{Step 3:} Using the linear programming models LP1 and LP2, determine the weight vectors based on kernel and greyness, denoted respectively by \[
	\boldsymbol{\omega}' = (\omega_1', \omega_2', \ldots, \omega_m') \quad \text{and} \quad
	\boldsymbol{\omega}'' = (\omega_1'', \omega_2'', \ldots, \omega_m'').
	\]
Then, compute the comprehensive weight vector as\ $
	\boldsymbol{\omega} = (\omega_1, \omega_2, \ldots, \omega_m).
	$	\\\textbf{Step 4:} Calculate the overall evaluation value for each alternative and perform ranking according to the obtained scores.
	\\
	Firstly, let us compute the overall weighted attribute value of $x_i$
	\[x_i=\bigoplus_{j=1}^m \sigma(a_j)\otimes \tilde r_{ij},\;\;
	where\;\; \hat x_i=\sum_{j=1}^m \hat w_j\,\hat{\tilde r}_{ij},\;\;
	g_{x_i}=\max_j\{g_{w_j}\vee g_{\tilde r_{ij}}\};\]
	\[
\tilde r_{ij}\;=\;\bigoplus_{p=1}^m \xi_{pj}\otimes r_{ip},\;\;where\;\;\hat{\tilde r}_{ij}=\sum_{p=1}^m \hat\xi_{pj}\hat r_{ip},\;
g_{\tilde r_{ij}}=\max_{p}\{g_{\xi_{pj}}\vee g_{r_{ip}}\}.
\]\\
Secondly, rank alternatives using the \emph{relative kernel}, sorting in descending order of $\delta_i$ (ties broken by larger $\gamma_i$)
	\[
	\gamma_i=\frac{1}{1+g_{x_i}},\;\;
	\delta_i=\gamma_i\,\hat x_i=\frac{\hat x_i}{1+g_{x_i}}.
	\]
\subsection{The Properties of This Method}
(1) The method is consistent with the axioms of grey arithmetic; 

(2) The influence network reduces to classical weighted aggregation when $\xi=I$; 

(3) Under the strong-graph option, edge kernels cannot exceed their endpoints, preventing over-confident relations; 

(4) All operations are closed in $[0,1]\times[0,1]$, and the greyness never decreases during aggregation.
\section{Illustrative Example and Analysis}
\subsection{Background and Scenario Description}
To demonstrate the practicality of the interval grey numbers multi-attribute decision-making model based on grey graph, a case concerning the selection of an optimal service system is presented.  
Assume that an organization is evaluating three potential service solutions, denoted as $X_1$, $X_2$, and $X_3$, which differ in \textit{cost}, \textit{performance}, and \textit{service quality}. Because of uncertainty in expert judgment and data variability, the evaluations for each attribute are expressed as interval grey numbers.

The three evaluation attributes are defined as follows
\[
A_1:\text{cost (cost-type)}, \ 
A_2: \text{performance (benefit-type)}, \ 
A_3: \text{service quality (benefit-type)}.
\]
A smaller value of $A_1$ indicates lower cost, while higher values of $A_2$ and $A_3$ imply better performance and service quality.

(1) The interval grey numbers decision matrix in Table \ref{tab:1} is therefore constructed as
\begin{table}[htbp]
	\centering
	\caption{Interval grey numbers decision making matrix.}
	\begin{tabular}{cccc}
		\toprule
		& $A_1$    & $A_2$     & $A_3$ \\
		\midrule
		$X_1$    & [90,110] & [70,85] & [60,75] \\
		$X_2$    & [80,95] & [65,80] & [70,85] \\
		$X_3$    & [85,100] & [80,90] & [55,70] \\
		\bottomrule
	\end{tabular}%
	\label{tab:1}%
\end{table}%

(2) The initial attribute weights, reflecting expert consensus, are defined as
\[
w_1 \in [0.40, 0.50], \quad w_2 \in [0.30, 0.40], \quad w_3 \in [0.15, 0.25].
\]
After normalization based on kernel–greyness unification, the kernel and greyness components are
\[
(\hat{w}_1, g_{w_1}) = (0.45, 0.10), \ 
(\hat{w}_2, g_{w_2}) = (0.35, 0.10), \ 
(\hat{w}_3, g_{w_3}) = (0.20, 0.10).
\]

\begin{figure}[H]
	\centering
	\includegraphics[width=0.3\linewidth]{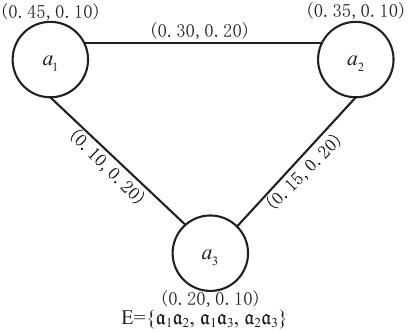}
	\caption{ Grey graph of attributes.}
	\label{fig:4}
\end{figure}

(3) The attribute influence relationship matrix, reflecting interdependencies among cost, performance, and service quality, is given as
\[
\hat{\xi} =
\begin{bmatrix}
	1 & 0.3 & 0.1 \\
	0.3 & 1 & 0.15 \\
	0.1 & 0.15 & 1
\end{bmatrix},
\ 
G_{\xi} =
\begin{bmatrix}
	0 & 0.2 & 0.2 \\
	0.2 & 0 & 0.2 \\
	0.2 & 0.2 & 0
\end{bmatrix}.
\]
Figure \ref{fig:4} illustrates the relationship between the above attribute.
\subsection{Computational Procedure}
The computational process follows the methodology outlined in the previous section.
\\
\textbf{Step 1:} Interval Normalization and Kernel–Greyness Conversion. The interval data are first normalized to $[0,1]$, distinguishing between cost-type and benefit-type attributes.  
The kernel and greyness of the normalized values are then computed as
\[
\hat{r} = \frac{r^L + r^U}{2}, \  g = r^U - r^L.
\]
After computation, the normalized decision matrix is then expressed as
\[
R = 
\begin{bmatrix}
	[0.0000,0.6667] & [0.2000,0.8000] & [0.1667,0.6667] \\
	[0.5000,1.0000] & [0.0000,0.6000] & [0.5000,1.0000] \\
	[0.3333,0.8333] & [0.6000,1.0000] & [0.0000,0.5000]
\end{bmatrix}.
\]
The normalized kernel and greyness matrices are
\[
\hat{R} =
\begin{bmatrix}
	0.3333 & 0.5000 & 0.4167 \\
	0.7500 & 0.3000 & 0.7500 \\
	0.5833 & 0.8000 & 0.2500
\end{bmatrix},
\ 
G =
\begin{bmatrix}
	0.6667 & 0.6000 & 0.5000 \\
	0.5000 & 0.6000 & 0.5000 \\
	0.5000 & 0.4000 & 0.5000
\end{bmatrix}.
\]
\\
\textbf{Step 2:} Attribute Influence Propagation. To incorporate the interdependence among attributes, the grey matrix multiplication method is applied
	\[
\tilde r_{ij}\;=\;\bigoplus_{p=1}^m \xi_{pj}\otimes r_{ip},\;\;where\;\;\hat{\tilde r}_{ij}=\sum_{p=1}^m \hat\xi_{pj}\hat r_{ip},\;\ 
g_{\tilde r_{ij}}=\max_{p}\{g_{\xi_{pj}}\vee g_{r_{ip}}\}.
\]
This yields the value of the interdependence among attributes
\[
\hat{\tilde r}_{ij} =
\begin{bmatrix}
	0.5250 & 0.6625 & 0.5250 \\
	0.9150 & 0.6375 & 0.8700 \\
	0.8483 & 1.0125 & 0.4283
\end{bmatrix},
\ 
g_{\tilde r_{ij}} =
\begin{bmatrix}
	0.6667 & 0.6667 & 0.6667 \\
	0.6000 & 0.6000 & 0.6000 \\
	0.5000 & 0.5000 & 0.5000
\end{bmatrix}.
\]
\\
\textbf{Step 3:} Weighted Aggregation of Grey Attributes. Each alternative’s comprehensive evaluation value is computed via
	\[x_i=\bigoplus_{j=1}^m \sigma(a_j)\otimes \tilde r_{ij},\;\;
where\;\; \hat x_i=\sum_{j=1}^m \hat w_j\,\hat{\tilde r}_{ij},\;\;
g_{x_i}=\max_j\{g_{w_j}\vee g_{\tilde r_{ij}}\}.\]
The overall weighted attribute values of each scheme are
\[
\begin{aligned}
	\hat{x}_1 &= 0.45\times0.5250 + 0.35\times0.6625 + 0.20\times0.5250 = 0.5731, \  g_{x_1}=0.6667, \\
	\hat{x}_2 &= 0.8089, \; g_{x_2}=0.6000, \\
	\hat{x}_3 &= 0.8218, \; g_{x_3}=0.5000.
\end{aligned}
\]
\\
\textbf{Step 4:} Ranking by Relative Kernel Score. The relative kernel and precision in Table \ref{tab:2} are obtained through
$
\gamma_i = \frac{1}{1 + g_{x_i}}, \ 
\delta_i = \frac{\hat{x}_i}{1 + g_{x_i}}.
$\begin{table}[H]
	\centering
	\caption{The score of relative kernel.}
	\begin{tabular}{cccc}
		\toprule
		& $\hat{x}_i$    & $g_{x_i}$     & $\delta_i = \dfrac{\hat{x}_i}{1+g_{x_i}}$ \\
		\midrule
		$X_1$ & 0.5731 & 0.6667 & 0.3439 \\
		$X_2$ & 0.8089 & 0.6000 & 0.5056 \\
		$X_3$ & 0.8218 & 0.5000 & 0.5479\\
		\bottomrule
	\end{tabular}%
	\label{tab:2}%
\end{table}
Based on the above calculation values, the ranking results of each alternative are as follows\[
X_3 \succ X_2 \succ X_1.
\]
\subsection{Result Discussion and Conclusion}
From the above analysis, $X_3$ achieves the highest relative kernel score, followed by $X_2$ and then $X_1$.  
This indicates that, while $X_3$ has moderate cost and superior performance, its balance between service quality and uncertainty (low greyness) makes it the most favorable option overall.  
In contrast, $X_1$ shows higher cost and higher uncertainty in performance, leading to a lower composite score.

In summary, the model provides a feasible and interpretable framework for uncertain decision-making environments. In this case, the final ranking $X_3 \succ X_2 \succ X_1$ indicates that the third service scheme provides the optimal balance between cost, performance, and service quality under uncertainty.
\section{Conclusion}
This example demonstrates the applicability and robustness of the grey graph multi-attribute decision-making model. The approach captures uncertainty in both evaluation and weighting, reduces information distortion during normalization, and ensures comparability across attributes. Compared with conventional single-valued decision models, the kernel–greyness and graph framework better reflects the decision-maker’s cognitive uncertainty and the internal relations among attributes. This case verifies that the proposed grey graph model effectively handles uncertain multi-attribute information. By simultaneously considering attribute interdependence, interval uncertainty, and decision-maker preferences, this model provides a structured and objective ranking mechanism suitable for real-world decision contexts such as technology selection, supplier evaluation, or service system optimization. This structure provides a flexible model for partially known relationships, such as those in social networks or environmental systems with measurement ambiguity.
\section*{Declaration of Competing Interest}
No conflicts of interest regarding the publication of this paper.
\section*{Data availability}
Data will be made available on request.




\bibliographystyle{elsarticle-num}
\bibliography{greybib}
\end{document}